\documentclass[aps,prl,article,twocolumn,preprintnumbers,amsmath,amssymb,superscriptaddress]{revtex4-1}

\date{\today}
\usepackage{epsfig}
\usepackage{subfigure}
\usepackage{graphicx}
\usepackage{dcolumn}
\usepackage{bm}
\usepackage[colorlinks,citecolor=blue,linkcolor=blue,hyperindex]{hyperref}
\usepackage{float}
\usepackage{hyperref}
\usepackage{comment}
\newcommand{\real}      {\mathrm{Re}}

\newcommand{\Ham}   {{\mathcal{H}}}

\newcommand{\qbf}      {{q}}

\newcommand{\ibf}      {\textbf{i}}

\newcommand{\rbf}      {\textbf{r}}

\newcommand{\Upl}{U_{\rm plrn}}

\begin{document}

\title{Phonon-Mediated Long-Range Attractive Interaction in One-Dimensional Cuprates}
\author{Yao Wang}\thanks{These two authors contributed equally to this work.}
\affiliation{Department of Physics and Astronomy, Clemson University, Clemson, South Carolina 29631, USA}
\author{Zhuoyu Chen}\thanks{These two authors contributed equally to this work.}
\affiliation{Geballe Laboratory for Advanced Materials, Stanford University, Stanford, California 94305, USA}
\affiliation{Stanford Institute for Materials and Energy Sciences, SLAC National Accelerator Laboratory, Menlo Park, California 94025, USA}
\affiliation{Departments of Physics and Applied Physics, Stanford University, Stanford, California 94305, USA}
\author{Tao Shi}
\affiliation{CAS Key Laboratory of Theoretical Physics, Institute of Theoretical Physics, Chinese Academy of Sciences, Beijing 100190, China}
\affiliation{CAS Center for Excellence in Topological Quantum Computation, University of Chinese Academy of Sciences, Beijing 100049, China}
\author{Brian Moritz}
\affiliation{Stanford Institute for Materials and Energy Sciences, SLAC National Accelerator Laboratory, Menlo Park, California 94025, USA}
\author{Zhi-Xun Shen}
\affiliation{Geballe Laboratory for Advanced Materials, Stanford University, Stanford, California 94305, USA}
\affiliation{Stanford Institute for Materials and Energy Sciences, SLAC National Accelerator Laboratory, Menlo Park, California 94025, USA}
\affiliation{Departments of Physics and Applied Physics, Stanford University, Stanford, California 94305, USA}
\author{Thomas P. Devereaux}
\email[All correspondence should be addressed to Y.W.(\href{mailto:yaowang@g.clemson.edu}{yaowang@g.clemson.edu}) and T.P.D. (\href{tpd@stanford.edu}{tpd@stanford.edu}) 
]{}
\affiliation{Geballe Laboratory for Advanced Materials, Stanford University, Stanford, California 94305, USA}
\affiliation{Stanford Institute for Materials and Energy Sciences, SLAC National Accelerator Laboratory, Menlo Park, California 94025, USA}
\affiliation{Department of Materials Science and Engineering, Stanford University, Stanford, California 94305, USA}

\date{\today}
 \begin{abstract}
Establishing a minimal microscopic model for cuprates is a key step towards the elucidation of a high-$T_c$ mechanism. By a quantitative comparison with a recent \emph{in situ} angle-resolved photoemission spectroscopy measurement in doped 1D cuprate chains, our simulation identifies a crucial contribution from long-range electron-phonon coupling beyond standard Hubbard models. Using reasonable ranges of coupling strengths and phonon energies, we obtain a strong attractive interaction between neighboring electrons, whose strength is comparable to experimental observations. Nonlocal couplings play a significant role in the mediation of neighboring interactions. Considering the structural and chemical similarity between 1D and 2D cuprate materials, this minimal model with long-range electron-phonon coupling will provide important new insights on cuprate high-$T_C$ superconductivity and related quantum phases.
 	\end{abstract}
\maketitle

The origin of high-$T_c$ superconductivity in cuprates remains one of the greatest mysteries in condensed matter physics\,\cite{bednorz1986possible,tsuei2000pairing,damascelli2003angle,armitage2010progress}. The microscopic mechanism is believed to be related to the strong correlations represented by the Hubbard model\,\cite{anderson1987resonating,zhang1988effective,eskes1988tendency}. Although numerical simulations using this model have reproduced some observations in cuprates, such as antiferromagnetism\,\cite{singh2002spin}, spin and charge stripe phases\,\cite{zheng2017stripe, huang2017numerical,huang2018stripe,ponsioen2019period}, and strange metallic behavior\,\cite{kokalj2017bad, huang2019strange, cha2020slope}, the most significant phase --- high-$T_c$ $d$-wave superconductivity --- remains an enigma. To date, numerical evidence for quasi-long-range-ordered superconductivity has been reported for specific systems and methods\,\cite{maier2005systematic, zheng2016ground,ido2018competition,jiang2019superconductivity,jiang2020ground}, but the exact solutions with cylinder geometry always reveal a coexistence of the charge order with comparable strength, and superconducting correlations progressively decay on shorter length scales as the numerical cluster size increases\,\cite{chung2020plaquette,jiang2021ground,gong2021robust,jiang2021stripe}. This contrasts sharply to the robust high-$T_c$ superconducting phases observed in a large family of cuprate compounds. Increasing experimental evidence in the low-energy regime, {\it e.g.} polaronic dressing near half-filling\,\cite{shen2004missing} and lattice dressing effects that manifest as kinks or replica bands in photoemission measurements\,\cite{lanzara2001evidence, tallon2005isotope,cuk2005review,lee2006interplay, reznik2006electron, devereaux2007inelastic,he2018rapid}, has suggested that small ingredients missing from the Hubbard model may have an outsized impact that can dramatically tip the balance towards some instability.

\begin{figure}[!t]
\begin{center}
\includegraphics[width=\columnwidth]{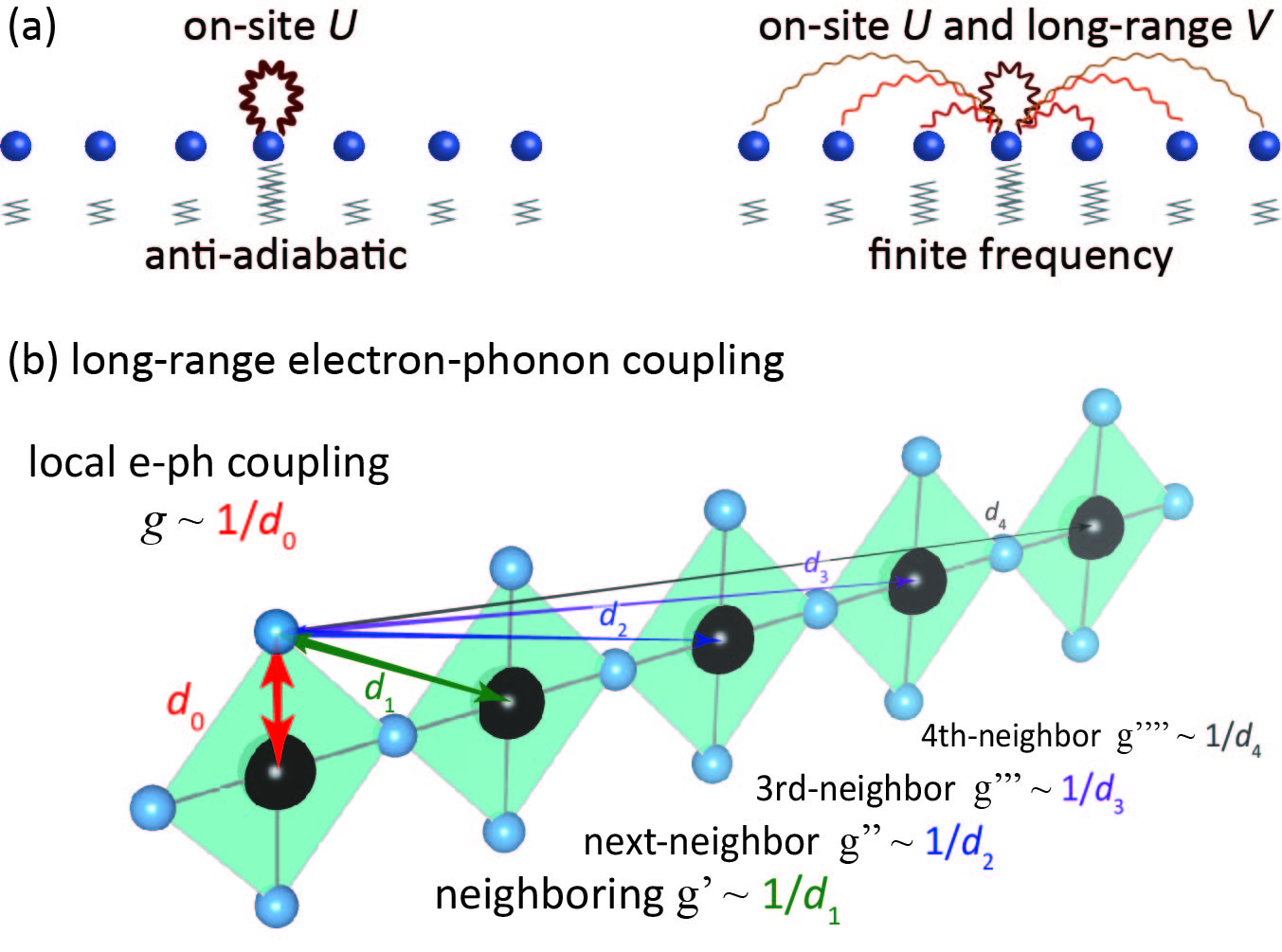}\vspace{-2mm}
\caption{\label{fig:cartoon} (a) Schematic diagram of a local effective interaction mediated by high-frequency phonons (antiadiabatic limit) and the nonlocality for finite frequencies. (b) Schematic explaining the geometry regarding local and nonlocal e-ph couplings, estimated by the inverse distance between the apical oxygen and corresponding copper atoms.\vspace{-6mm}
}
\end{center}
\end{figure}

To understand models with multiple degrees of freedom, which all play a significant role at low energies, presents technical challenges in dealing with the coexistence of strong correlations and quantum fluctuations\,\cite{dagotto1994correlated,keimer2015quantum}, intertwined instabilities, and microscopic competition\,\cite{davis2013concepts,fradkin2015colloquium}.  Theoretical calculations in 2D are limited by the rapid increase of Hilbert-space dimension and entanglement with system size. Hubbard-like correlated models in 2D, and in the thermodynamic limit, have yielded only limited rigorous results. An alternative approach to better understand the problem may lie in an examination of the microscopic ingredients necessary to describe 1D cuprate analogs with comparable structural and chemical environments\,\cite{fujisawa1999angle, kim2006distinct, schlappa2012spin}.  

With better control of theory in 1D systems, quantitative comparisons to experiment can be made with higher fidelity, which enable a determination of the most significant missing microscopic ingredients. Recently, \emph{in situ} molecular beam epitaxy and angle-resolved photoemission techniques have enabled a study of the single-particle spectral function across a range of doping in 1D cuprate chains\,\cite{chen2020anomalously}.  These experiments have revealed an anomalously strong ``holon folding'' near $k_F$ with the same velocity as the holon, which reflects interactions in the charge channel. The intensity of this ``holon folding'' spectral feature cannot be captured in simulations of the typical single-band Hubbard model. Only by adding a substantial, attractive, near-neighbor interaction to the model can the theoretical simulations well explain the experimental observations\,\cite{chen2020anomalously}. Because of the repulsive nature of the electrostatic interaction between electrons, such an attraction must be mediated by a virtual process that may involve degrees of freedom not present in the single-band electronic model. This interaction is anomalously strong, far exceeding the effective $t^2/U$ density interaction term obtained by a Schrieffer-Wolff transformation of the Hubbard model. Motivated by existing experimental evidence for strong lattice effects in cuprates, a possible explanation may lie in the coupling between electrons and phonons (e-ph), although this still lacks a quantitative demonstration.

Here, we demonstrate that the experimentally observed strong attractive interaction $V$ can be addressed naturally in a model that includes a reasonable, long-range e-ph coupling, determined from Madelung potentials and Franck-Condon fitting\,\cite{lee2013role}. An intuitive argument is sketched in Fig.~\ref{fig:cartoon}, where retardation effects spread the effective attraction mediated by phonons over longer distances.  While retardation itself gives rise to a weak, near-neighbor, attractive interaction in the Hubbard-Holstein model, we find that a long-ranged e-ph coupling determined by the lattice geometry [{\it i.e.}~$g^\prime$, $g^{\prime\prime}$, and $g^{\prime\prime\prime}$ in Fig.~\ref{fig:cartoon}(b)] provides a substantial enhancement necessary to account for the experimental observations. Considering the structural and chemical similarity between these 1D chain and 2D planar cuprate materials, a combination of the well-known electron correlations present in the Hubbard model and long-ranged e-ph coupling provides a minimal description on which to base future cuprate studies.

\begin{figure*}[!t]
\begin{center}
\includegraphics[width=18cm]{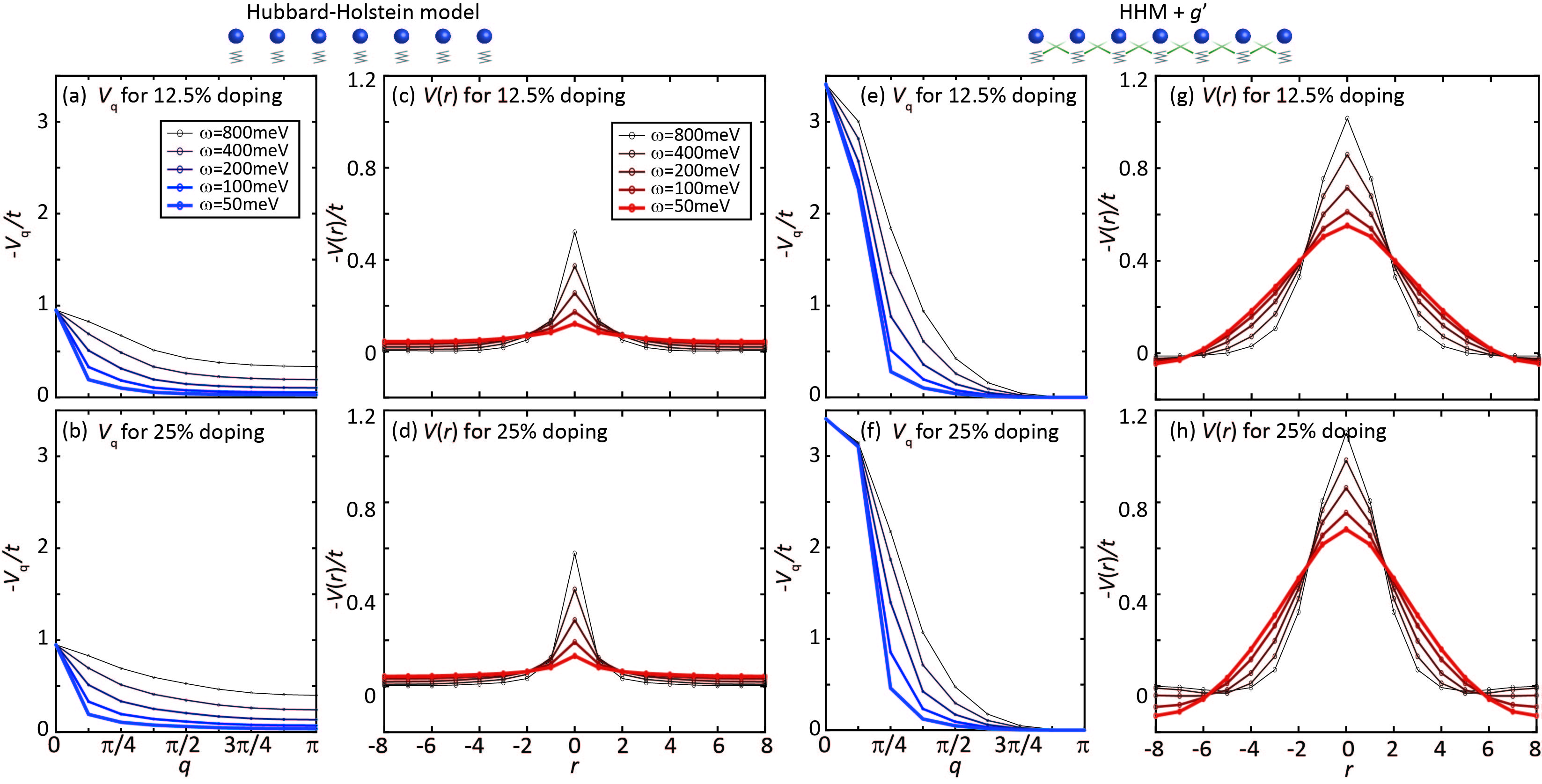}\vspace{-2mm}
\caption{\label{fig:HHMvsgpr} Effective interactions obtained from the Hubbard-Holstein model (left) and the Hubbard-extended-Holstein model (HHM+$g^\prime$, right). (a) The momentum distribution of $V_\qbf$ as a function of $q$, for different phonon frequencies with fixed $\lambda = 0.95$. (c) The spatial distribution of $V(r)$ as a function of distance $r$. Both (a) and (c) are obtained for 12.5\% doped HHM. (b,d) Same as (a,c) but for 25\% doped HHM. (e-h) Same as (a-d) but for the Hubbard-extended-Holstein model with $g^\prime=g/\sqrt{5}$.\vspace{-4mm}
}
\end{center}
\end{figure*}

We consider a Hubbard-Holstein-like model with strong on-site Coulomb repulsion
\begin{eqnarray}\label{eq:HHM}
    H&=&-\sum_{i\sigma} t (c^\dagger_{i\sigma}c_{i+1,\sigma} + H.c.) + U\sum_i n_{i,\uparrow}n_{i,\downarrow} \nonumber\\
    &&+ \sum_{i,j,\sigma} g_{ij} n_{i\sigma} (a_j^\dagger + a_j) + \sum_i \omega a_i^\dagger a_i
\end{eqnarray}
where $c_{i\sigma}$ ($c_{i\sigma}^\dagger$) annihilates (creates) an electron at site $i$ with spin $\sigma$ and $a_j$ ($a_{j}^\dagger$) annihilates (creates) a phonon at site $j$. The bare electronic kinetic and potential energies are parameterized, respectively, by the hopping integral $t$, and the on-site Coulomb interaction $U$, the largest energy scale in this microscopic model. We ignore the extended Coulomb interactions originating from electronic repulsions, as they are substantially screened by the copper-oxygen covalent bond. The phonons are treated as Einstein optical modes with bare frequency $\omega$, and a real-space coupling $g_{ij}$ between the charge density $n_{i}$ on site $i$ and phonon displacement on site $j$. A local restriction on the e-ph coupling $g_{ij}$ ({\it i.e.}~$g_{ij}=g\,\delta_{ij}$) reduces Eq.~\eqref{eq:HHM} to the standard Hubbard-Holstein model (HHM). With translation symmetry, periodic boundary conditions, and no disorder, $g_{ij}$ can be expressed as a function of only the relative distance between sites $|i-j|$; and it is convenient to express this in reciprocal space (bosonic momentum) as $g_{\qbf}$. In this latter representation, a momentum independent $g_{\qbf}$ corresponds to a spatially local coupling, while a strong momentum dependence indicates a longer-range coupling.

The physical properties of the 1D HHM have been studied by various numerical methods such as exact diagonalization (ED)\,\cite{hotta1997unconventional,fehske2002peierls,fehske2003nature,fehske2004quantum}, density-matrix renormalization group (DMRG)\,\cite{tezuka2005density,tezuka2007phase,fehske2008metallicity,ejima2010dmrg}, and quantum Monte Carlo (QMC)\,\cite{clay2005intermediate,hardikar2007phase,hohenadler2013excitation,greitemann2015finite}. Here, we employ a recently developed variational non-Gaussian ED (NGSED) method\,\cite{wang2020zero}, which has been benchmarked with exact QMC results for the HHM\,\cite{wang2020zero,costa2020phase} and can be extended easily to longer-range interactions like those considered in Eq.~\eqref{eq:HHM}. More importantly, NGSED provides direct information about the phonon mediated effective interations $V$, which can be used to benchmark  parameters extracted from experimental comparisons\,\cite{chen2020anomalously}. Following the NGSED framework, we consider a wavefunction ansatz\,\cite{wang2020zero}
\begin{eqnarray}\label{eq:wvfuncansatzFull}
    \big|\Psi\big\rangle  &=& \Upl(f_\qbf) |\psi_{\rm ph}\rangle \otimes|\psi_{\rm e}\rangle, \\
    \Upl(f_\qbf) &=& e^{i\frac1{\sqrt{N}}\sum_\qbf f_\qbf p_{-\qbf}. \rho_\qbf},
\end{eqnarray}
as the solution to this strongly correlated model. Here, the momentum-space electron density $\rho_{\qbf}=\sum_{\ibf\sigma} n_{\ibf\sigma}e^{-i\qbf\cdot \rbf_\ibf}$ and phonon momentum operator $p_{\qbf}=i\sum_{\ibf}( a_{\ibf}^{\dagger}-a_{\ibf})  e^{-i\qbf\cdot \rbf_\ibf} /\sqrt{N}$. The right-hand side is a direct product of electron and phonon states (denoted as $|\psi_{\rm e}\rangle$ and $|\psi_{\rm ph}\rangle$, respectively), where $|\psi_{\rm e}\rangle$ is treated as a full many-body state while $|\psi_{\rm ph}\rangle$ is a Gaussian state\,\cite{shi2018variational}. The polaron transformation $\Upl(f_\qbf)$ in Eq.~\eqref{eq:wvfuncansatzFull} entangles the two parts of the wavefunction. Note that different from the Lang-Firsov transformation\,\cite{lang1962}, the $f_\qbf$'s are momentum-dependent variational parameters determined self-consistently within NGSED\,\cite{wang2020zero}.

Within the wavefunction ansatz of Eq.~\eqref{eq:wvfuncansatzFull}, the ground state is obtained by iteratively minimizing the energy $E =  \big\langle\Psi\big|\Ham \big|\Psi\big\rangle$ with respect to the electronic wavefunction $|\psi_e\rangle$ and the variational parameters. With fixed variational parameters, the effective electronic Hamiltonian is given by tracing over the phonon state\,\cite{wang2020zero}
\begin{eqnarray}
    \Ham^{\rm (eff)}_{e} = \langle\psi_{\rm ph} | \Upl(f_\qbf)^\dagger \Ham \Upl(f_\qbf)|\psi_{\rm ph}\rangle\,.
\end{eqnarray}
The polaronic dressing is reflected in the electronic Hamiltonian $\Ham^{\rm (eff)}_{e}$ through its effective kinetic energy and the additional electronic attraction mediated by phonons [for a detailed derivation see Ref.~\onlinecite{wang2020zero} or the Supplementary Material \cite{SUPMAT}]
\begin{eqnarray}\label{eq:effInteraction}
    V_\qbf = 2 \omega|f_\qbf|^2 -4g_\qbf \real[f_\qbf].
\end{eqnarray}
Physically, this effective $V_\qbf$ is sketched in Fig.~\ref{fig:cartoon}(a): antiadiabatic phonons ($\omega\rightarrow\infty$) can be integrated out and lead to a closed-form interaction $V_\qbf\equiv -2g_q^2/\omega$. For a Holstein-type coupling ($g_q=g$), this (antiadiabatic) interaction is momentum independent, or equivalently, local in coordinate space. It implies that the lattice potential mirrors the instantaneous variation of the local electron density and immediately acts back on that density. The net effect is a correction to the on-site Coulomb interaction. However, at finite frequency the phonons are retarded and carry information about the electron hopping, mediating a nonlocal interaction [see Fig.~\ref{fig:cartoon}(a)]. Such a nonlocal effect already will be present with an infinitesimally small e-ph coupling [see the Supplementary Materials \cite{SUPMAT} for discussions].

\begin{figure*}[!t]
\begin{center}
\includegraphics[width=18cm]{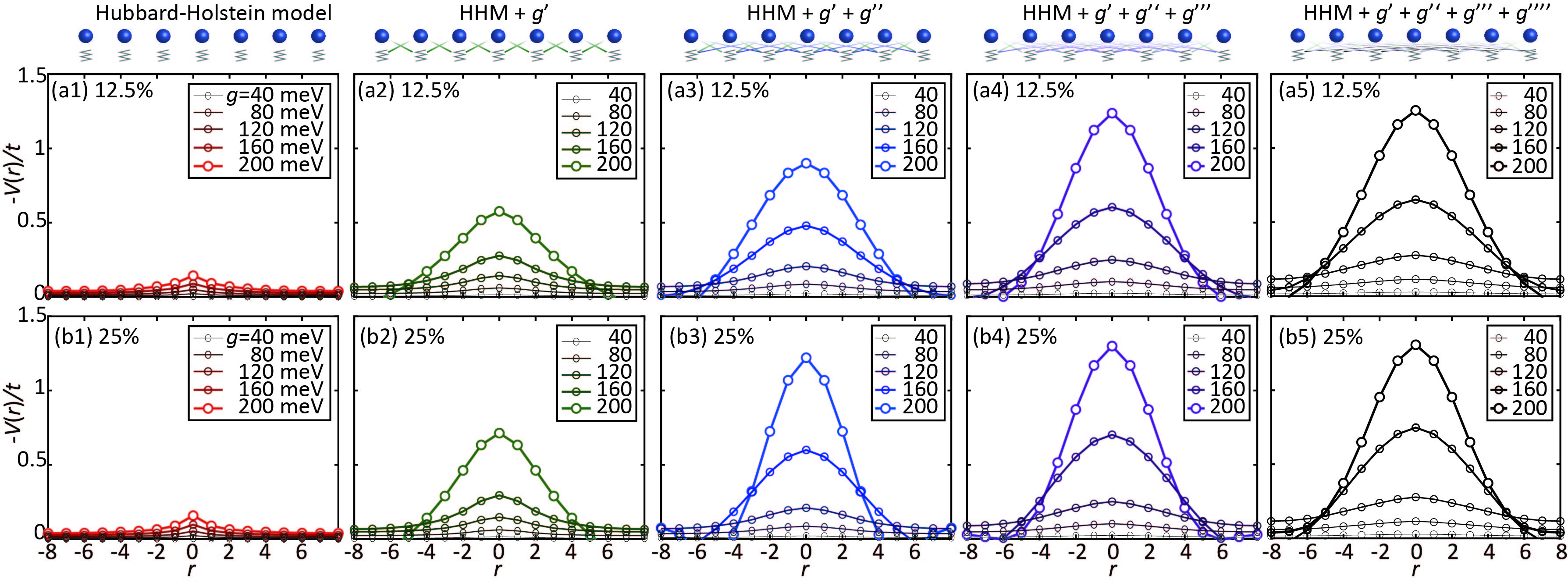}\vspace{-4mm}
\caption{\label{fig:fullVrCompare} Comparison of the effective $V(r)$ mediated by e-ph coupling of different ranges. (a1-a5) The spatial distribution of $V(r)$ as a function of distance $r$ for a 12.5\% doped (a1) HHM, (a2) 12.5\% doped HHM+$g^\prime$ (with $g^\prime=g/\sqrt{5}$), (a3) HHM+$g^\prime$+$g^{\prime\prime}$ (with $g^{\prime\prime}=g/\sqrt{17}$), (a4) HHM+$g^\prime$+$g^{\prime\prime}$+$g^{\prime\prime\prime}$ (with $g^{\prime\prime\prime}=g/\sqrt{37}$), and (a5) HHM+$g^\prime$+$g^{\prime\prime}$+$g^{\prime\prime\prime}$+$g^{\prime\prime\prime\prime}$ (with $g^{\prime\prime\prime\prime}=g/\sqrt{65}$). The phonon frequency is fixed at $\omega=70$\,meV.  (b1-b5) Same as (a1-a5) but for 25\% doping.\vspace{-4mm}
}
\end{center}
\end{figure*}

We first consider the HHM with only local e-ph coupling at an intermediate strength $\lambda=g^2/\omega=0.95$ (such a value, while serving an illustrative purpose here, will be justified on physical grounds later), using a 16-site chain with periodic boundary. After self-consistently solving for the ground-state wavefunctions using NGSED, we obtain the phonon-mediated attraction $V_q$ shown in Figs.~\ref{fig:HHMvsgpr}(a) and (b). With an increasing phonon frequency, $V_q$ exhibits weaker momentum dependence. To quantify the neighboring attractive interaction, we extract the spatial distribution of $V(r)$, which decreases rapidly with distance, resulting in a near-neighbor interaction $\sim0.1t$. The comparison between experiment and theory presented in Ref.~\onlinecite{chen2020anomalously} suggests that a near-neighbor attraction $\sim t$ is required to account for the observed holon folding\,\cite{chen2020anomalously}, while the attractive interactions presented in Figs.~\ref{fig:HHMvsgpr}(a) and (b) are an order of magnitude smaller than that experimental assessment. Therefore, the results from a pure HHM fail to provide a strong $V_{\rm eff}=V(r=1)$ unless we increase the coupling to an unphysically large strength. 

To address the effective interactions within a reasonable range of e-ph coupling, we note that the electrostatic interaction has an intrinsically long-range nature. As shown in Fig.~\ref{fig:cartoon}(b), for the case where the phonon originates from the motion of the apical oxygens, the e-ph interaction strengths at different sites can be approximated through the geometric distances. The local e-ph coupling strength is proportional to $1/d_0\sim 2/a$, where $a$ is the lattice constant for a CuO unit cell. The ratios for subsequent Cu-O distances can be read off immediately: $d_1/d_0 \approx \sqrt{5}$, $d_2/d_0 \approx \sqrt{17}$, and $d_3/d_0 \approx \sqrt{37}$. The influence on the electrons in the CuO chain can be approximated by the potential proportional to $1/d_n$, so $g^\prime = g/\sqrt{5}$, $g^{\prime\prime} = g/\sqrt{17}$, and $g^{\prime\prime\prime} = g/\sqrt{37}$. (Here, we employ the simplest geometric relation to give an order-of-magnitude estimation. A more realistic model should also consider the anisotropy, the dielectric constant, and the integration towards the thermodynamic limit\,\cite{lee2014interfacial}.)

The impact of these long-range e-ph couplings is prominent. Figures~\ref{fig:HHMvsgpr}(e-h) present the momentum and spatial distribution of $V_\qbf$, including the influence of $g^\prime$, for different phonon frequencies and a fixed $\lambda=0.95$ (i.e., all coupling strengths scale quadratically with the frequencies). Compared to the local (Holstein) coupling in Figs.~\ref{fig:HHMvsgpr}(a-d), this effective interaction with nonlocal $g^\prime$ produces a more momentum dependent $V_q$, which is further amplified for small phonon frequencies. In real space, this $q$ dependence corresponds to a nonlocal $V(r)$. Therefore, the near-neighbor effective interaction $V_{\rm eff}$ obtained in the Hubbard-extended-Holstein model increases for two reasons: (1) the additional $g^\prime$ increases the strength of the interaction mediated by phonons; (2) $g^\prime$ provides a direct, nonlocal interaction between electrons and phonons. The effective $V(r)$ for both 12.5\% and 25\% doping have comparable values, as it describes how phonons dress neighboring electrons and mediate local attraction, with little relevance to other electrons. This observation is consistent with experiments, in the sense that independent fittings of all spectral functions for different doping levels result in an almost uniform value of the near-neighbor, attractive interaction $V=-t$\,\cite{chen2020anomalously}, which lends further support to the ideas presented here that phonons are ultimately responsible for this attraction.

\begin{figure}[!t]
\begin{center}
\includegraphics[width=\columnwidth]{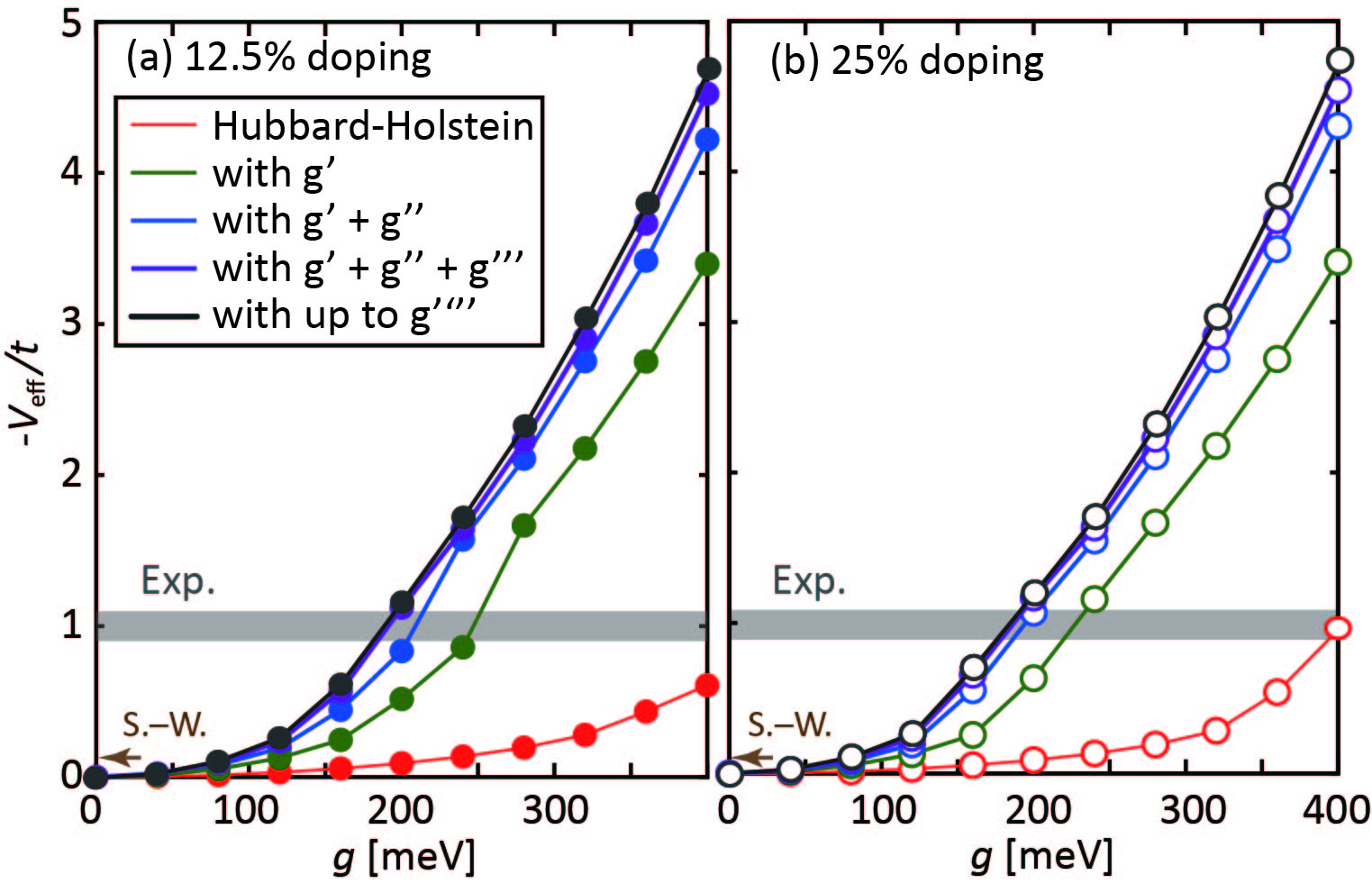}
\caption{\label{fig:extractedVvalues} The effective neighboring interaction $V_{\rm eff}$ extracted from the variants of models for (a) 12.5\% doping and (b) 25\% doping. Different colors denote the longest range of e-ph coupling included in the model: local e-ph coupling $g$ (red), up to neighboring e-ph coupling $g^\prime$ (green), up to next-, third-, and fourth-neighboring e-ph coupling $g^{\prime\prime}$ (blue, purple, and black). For different $g$ values, the ratios between allowed nonlocal couplings are fixed as the geometric ratios in Fig.~\ref{fig:cartoon}(b). The gray line denotes the experimentally extracted strength, while the brown arrow denotes the $V_{\rm eff}$ obtained by the Schrieffer-Wolff transformation of the Hubbard model.\vspace{-4mm}
}
\end{center}
\end{figure}

Armed with knowledge of the significant effects of nonlocal e-ph couplings, we now investigate the impact of coupling strengths and the range of the interaction. To provide a basis for realistic comparison, we fix $\omega=$70\,meV in the following discussion, identified in cuprates with CuO$_2$ plaquette modes\,\cite{rosner1997electronic,johnston2010systematic, lee2013role}. As shown in Figs.~\ref{fig:fullVrCompare}(a1) and (b1), the interaction strength extracted from the HHM with only local e-ph coupling leads to a small $V_{\rm eff}$, impractical if the goal were to reach the experimentally observed value  $\sim t$ for reasonable coupling $g$\,\cite{chen2020anomalously}. However, introducing $g^\prime$, even restricted by geometric considerations to $g^\prime = g/\sqrt{5}$, we find that $V_{\rm eff}$ increases rapidly with $g$ [see Figs.~\ref{fig:fullVrCompare}(a2) and (b2)]. A coupling $g\sim200$\,meV (i.e.~$\lambda\sim0.95$) can produce $V_{\rm eff}\sim-0.5t$. More encouragingly, the addition of next-nearest-neighbor coupling $g^{\prime\prime}=g/\sqrt{17}$ further increases $V_{\rm eff}\sim-t$ for the same local coupling strength [see Figs.~\ref{fig:fullVrCompare}(a3) and (b3)]. To address the full impact of long-range e-ph coupling on the effective electrostatic interaction, we include also $g^{\prime\prime\prime}$ and $g^{\prime\prime\prime\prime}$, finding that $V_{\rm eff}$ saturates asymptotically [see Figs.~\ref{fig:fullVrCompare}(a4-a5) and Figs.~(b4-b5)]. Although these couplings still increase $V_{\rm eff}$, their impact is not as evident, as it is $g^\prime$ that connects neighboring electrons and phonons and has an outsized impact on the nearest-neighbor interaction. 

To better visualize the influence of e-ph coupling with different ranges and strengths, we extract the $V_{\rm eff}$ from Fig.~\ref{fig:fullVrCompare} and plot the dependence on $g$ in Fig.~\ref{fig:extractedVvalues}. With the asymptotically converged $V_{\rm eff}$, involving long-range coupling effects, we conclude that realistic e-ph coupling corresponding to experimental observations, and falling within the scenario presented here, should be $g \lesssim 185$\,meV.  Such a value is consistent with estimates from Madelung potential calculations and Franck-Condon fitting in another quasi-1D cuprate compound\,\cite{lee2013role}, suggesting that long-range e-ph can account adequately for the anomalously strong near-neighbor attraction derived from recent experiments\,\cite{chen2020anomalously}. If one further considers the extended Coulomb interaction arising from electronic repulsion, the total $V_{\rm eff}$ may be corrected slightly by $\sim$0.2$t$\,\cite{wohlfeld2010t}, within the error bar of experiments. Taking the parameters $t=600$\,meV and $\omega=70$\,meV extracted from experiments, this e-ph coupling corresponds to $\lambda = 0.81$, which is of moderate strength in correlated materials. This phonon-mediated $V_{\rm eff}$ is an order of magnitude stronger than that originated from the Schrieffer-Wolff transformation of the Hubbard model. Note that the {minimal} model adopted in the photoemission experiment \cite{chen2020anomalously} included only nearest-neighbor $V(r=1)$. If one were to consider even longer-range neighboring attraction $V(r>1)$, the corresponding $g$ to provide a good fit would be even smaller. 

To summarize, we conducted a systematic study of the Hubbard-extended-Holstein model and investigated the impact of phonon frequency and long-range e-ph coupling on the effective electronic attraction $V(r)$. Taking the e-ph coupling parameters extracted from existing studies of 1D cuprate materials, our simulation gives rise to an anomalously strong near-neighbor attractive interaction ($V\sim -t$), consistent with recent {\it in situ} angle-resolved photoemission experiments\,\cite{chen2020anomalously}. Our work has uncovered a significant missing ingredient in the microscopic description of 1D cuprate chains; and we have developed a minimal model that captures the essential experimental features. More generally, the similarities between 1D and 2D cuprates may be exploited to extend our conclusions to high-$T_c$ cuprate materials and the $d$-wave pairing, with limited corrections to model parameters. Future systematic DMRG and QMC studies are promising to extend the conclusion towards superconductivity.

Y.W. acknowledges support from the National Science Foundation (NSF) award DMR-2038011. Z.C., B.M., Z.X.S, and T.P.D. were supported by the Department of Energy, Office of Basic Energy Sciences, Materials Sciences and Engineering Division under Contract No. DE-AC02-76SF00515 for work at SLAC National Accelerator Laboratory and Stanford University. The calculations were performed at the Texas Advanced Computing Center (TACC).

\bibliography{paper}
\end{document}